# THE CONCEPT OF ISOCHORIC CENTRAL SPARK IGNITION AND ITS FUEL GAIN IN INERTIAL FUSION

A. Ghasemizad – M. Kamran

Physics department, Faculty of science, University of Guilan
P.O.BOX: 41335-1914, Rasht, Iran ( Islamic Republic of )
Corresponding author email: ghasemi@guilan.ac.ir

**ABSTRACT**
One of the best methods in inertial confinement fusion (ICF) is the concept of central spark ignition, consisting of two distinct regions named as hot and cold regions and formed by hydro-dynamical implosion of fuel micro-sphere central spark ignition method in inertial fusion and fuel pellet design condition in fusion power plant has been investigated and fuel gain for isochoric model in this method is calculated. We have shown the effects of different physical parameters of inertial fusion on fuel gain and optimized limit for fuel density and fuel pellet radius ($H_f = \rho R_f$) has been calculated.

## 1. Introduction

The essential condition for functioning a nuclear power plant that compression and ignition in it is done by laser beam is [1]:

$$\varepsilon_{He} f_e \varepsilon_{el} G_p > 1 \qquad (1)$$

where $\varepsilon_{He}$ is the transformation coefficient for deposited thermonuclear energy caused by nuclear fusion reaction to electrical energy, $f_e$ is the energy ratio of the exiting beam to the electric energy entering to the laser system, $\varepsilon_{el}$ is the transformation coefficient of entering electric energy to the electric energy getting to laser(or entering gian coefficient of the system), and $G_p$ is the energy multiplication coefficient in pellet (or pellet gain). For the minimum value of eq.(1) the values of $\varepsilon_{He}=0.4$, $f_e=0.25$, $\varepsilon_{el}=0.05$ and $G_p=200$ can be considered.
For calculation of energy multiplication coefficient the following equation can be used [2]:

$$G_p = \varepsilon_{lf} G_f \qquad (2)$$

where $\varepsilon_{lf}$ represents the beam energy fraction used for fuel compression inside the pellet and $G_f$ is fuel gain. A value for $\varepsilon_{lf}$ is 0.05[2]. In this article, fuel gain will be calculated and the effects of different physical parameters on it will be considered.

## 2. Description and circumstance of calculations

The isomole fuel of deuterium-tritium fuses and thermonuclear energy is released:

$$d + t \rightarrow n + \alpha + 17.6 MeV \qquad (3)$$

where the energies of alpha particle and neutron particles are 3.5 and 14.1 MeV respectively. We suppose all these energies to be deposited in fusion reactor core. If we consider neutron energy leakage from nuclear fusion environment, an outstanding loss in fuel gain is created that it is investigated in appendix. The fuel Pellets with initial fuel masses 1,3,5,7,10mg are considered and supposed to have two central region with radius $R_s$ and non-central region with radius $R_f$ ($R_s \ll R_f$). In isochoric model which fuel density in two central and non-central regions are equal, at first by a laser beam, we compress symmetrically the fuel inside the pellet and then by a laser beam with higher intensity, ignition in central region is created. Then the heat caused by this ignition enters the cold fuel around the central region and embraces the entire pellet. Therefore fuel gain of this system can be written as proportion of the deposited thermonuclear energy to the energy used for fuel compression and ignition:

$$G_f = \frac{E_f}{E_c + E_s} \qquad (4)$$

where, $E_f$, $E_c$ and $E_s$ are deposited thermonuclear energy, the energy used for fuel compression and the energy which causes the initial ignition respectively. In next section, these values will be calculated.

## 2-a. Calculation of thermonuclear energy

If $q_{dt}$ is the deposited energy from mass unite of reaction (3), $f_b$ is fuel fraction of used fuel inside pellet and $M_f$ is total fuel mass inside pellet, then $E_f$ can be calculated by:

$$E_f = q_{dt} \, f_b \, M_f \qquad (5)$$

For calculation of $f_b$, we have [3]:

$$f_b = \frac{H_f}{H_f + H_b} \quad , \quad H_f = \rho \, R_f \qquad (6)$$

$$H_b = \frac{8}{\sigma} \, M_i > 7 \text{gr/cm}^2 \qquad (7)$$

$$\sigma = \frac{1}{C_s} <\sigma v>_{dt} \qquad (8)$$

$$C_s = \left(\frac{3KT_i}{M_i}\right)^{1/2} \qquad (9)$$

where $\rho$, $M_i$, $T_i$, $K$, $<\sigma v>_{dt}$ and $C_s$ are fuel density, average ion mass, ion temperature, Boltzmann's constant, $\sigma - v$ parameter and sound speed in DT plasma. By substitution of equations (6) to (9) in eq.(5) for deposited thermonuclear energy we will have:

$$E_f(MJ) = 3.37 \times 10^5 \, M_f \, \frac{H_f}{H_f + 7} \qquad (10)$$

Where $M_f$ is in gr. In Figs. 1, 2 and 3 the variations of $E_f$ versus $M_f$ and $H_f$ are shown.

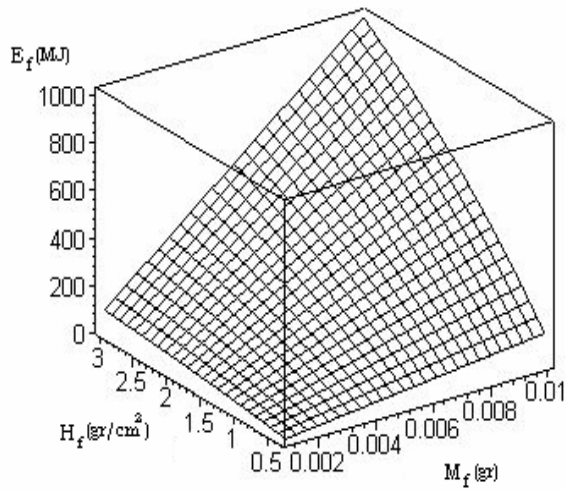

**Fig.1** The three dimensional variations of $E_f$ versus $H_f$ and $M_f$.

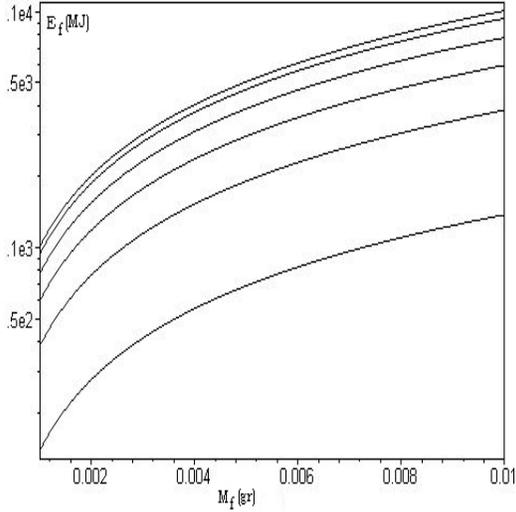 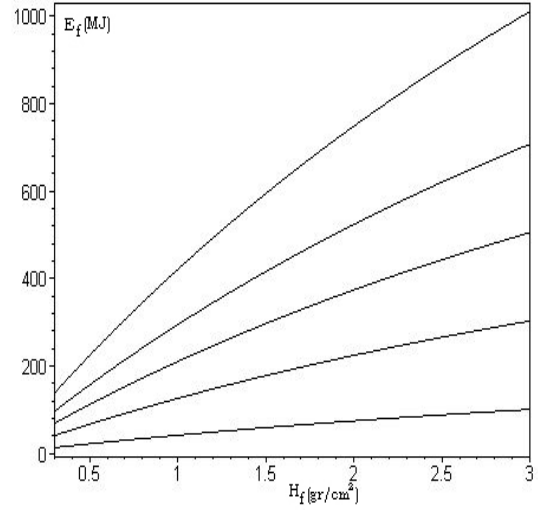

**Fig.2** The variations of $E_f$ versus $M_f$ for different values of $H_f$ from top 3, 2.7, 2.1, 1.9, 1.5, 0.9, .3 gr/cm$^2$

**Fig. 3** The variations of $E_f$ versus $H_f$ for Different values of $M_f$ from top 10, 7, 5, 3, 1 mgr

As Fig.1 shows, with increasing $M_f$ and $H_f$, deposited thermonuclear energy increases and the best case is the condition that in a pellet with a constant radius, more initial mass from fuel is placed, because in this condition fuel density and consequently $H_f$ increase. Fig.2 shows that in constant value of initial fuel mass, increasing of $H_f$ value causes that $E_f$ value to be increased and Fig.3 shows that in constant $H_f$ value, increasing of $M_f$ causes $E_f$ to be increased. In summary it can be yielded that the bigger fuel pellets with more fuel mass leads to the more created thermonuclear energy.

**2-b. Calculation of cold fuel compression energy**

The energy required for the compression of cold fuel, that its density is η time's more than liquid hydrogen density, can be calculated by [2]:

$$E_c = 0.12\, \alpha\, \eta^{2/3}\, M_f \quad , \quad \alpha > 1 \tag{11}$$

Where in this equation, α is called isentrope parameter and shows the deviation from fuel complete degeneracy [4]. In this article, we have done our calculations for α=1,2,3,4 and with the change of variables as follows, we obtain convenient form for $E_c$ :

$$H_f = H_0\, \eta^{2/3} \tag{12}$$

$$H_0 = (3\rho_0^2\, M_f / 4\pi)^{1/3} \tag{13}$$

Substitution of equations (12) and (13) in equation (11) yields:

$$E_c(MJ) = 0.12\, \alpha\, M_f\, H_f / (3\, \rho_0^2\, M_f / 4\pi)^{1/3} \tag{14}$$

The dependency of $E_c$ on isentrope parameter in above equation is clear and in Fig.4 we have shown that with increasing isentrope parameter, the energy needed for fuel compression increases. Also, Fig.5 shows that in constant values of initial fuel mass and η = 1500, increasing of isentrope parameter causes $E_c$ value to be increased. In summary it can be yielded that if we want the bigger fuel pellet or the more compressed fuel, the more compressing energy ( $E_c$ ) will be needed.

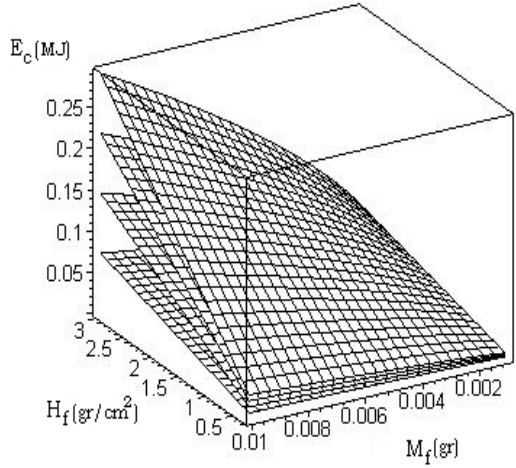 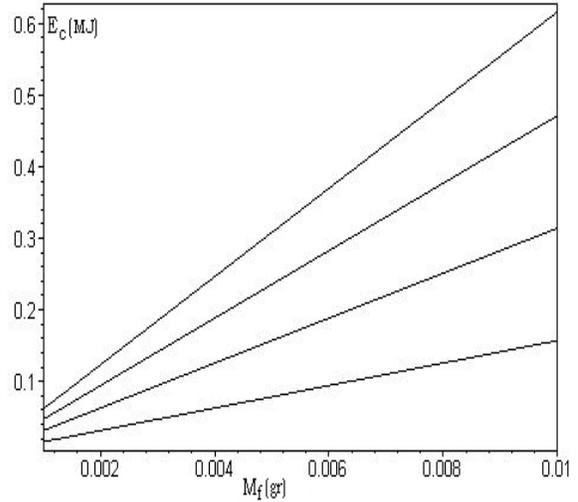

**Fig. 4** The three dimensional variations of $E_c$ versus $M_f$ and $H_f$ with different value of α parameter from top 4,3,2,1.

**Fig. 5** The variations of $E_c$ versus $M_f$ for η = 1500. and different values of α parameter from top 4,3,2,1.

## 2.c. Calculation of ignition energy

With introducing the ratio of fuel mass in hot central region to total fuel mass in the pellet with $f_s$, we can write:

$$f_s = (\frac{H_s}{H_f})^3 \qquad (15)$$

If $M_i$ is average ion mass and $T_s$ is ignition temperature in central region, for ignition energy we have [2]:

$$E_s = 3 f_s T_s (M_f / M_i) \qquad (16)$$

For example, for $T_s$=20KeV and $H_s$=0.3gr/cm², $E_s$ can be calculated with the following relation:

$$E_s(MJ) = 64.8 \, M_f / H_f^3 \qquad (17)$$

The dependency of $E_s$ on fuel mass and $H_f$ in above equation is clear and in fig.6 we have shown that in constant ignition temperature ( $T_s$ = 20 KeV ) and fuel mass ( $M_f$ ), increasing of $H_f$ value causes $E_s$ value to be increased. In summary it can be yielded that for bigger fuel pellet or the more fuel mass in pellet we will need the more spark energy.

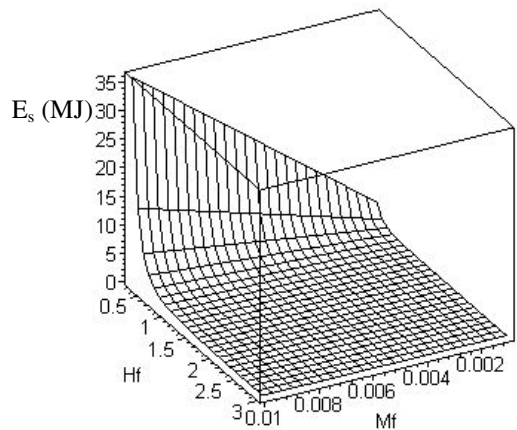

**Fig.6** The three dimensional variations of $E_s$ versus $M_f$ and $H_f$ in $T_s$=20 KeV.

## 2-d. Calculation of fuel gain

The substitution of equations (17),(14) and (10) in equation (11) yields fuel gain in parameters of $M_f$, $H_f$, $T_s$ and $\alpha$.

$$G_f = \frac{33700 \dfrac{M_f H_f}{H_f + 7}}{0.526\alpha H_f M_f^{2/3} + 0.00324 T_s \dfrac{M_f}{H_f^3}} \quad (18)$$

Generally $G_f$ is yielded from the following equation and figures 7, 8 and 9 show the circumstance of $G_f$ variations for these parameters. Fig. 5 shows the fuel gain, $G_f$ variations in $H_f$ and $M_f$ parameters. As we have shown in Fig. 7 for a constant value of $H_f$, more fuel mass inside the pellet yields higher fuel gain.

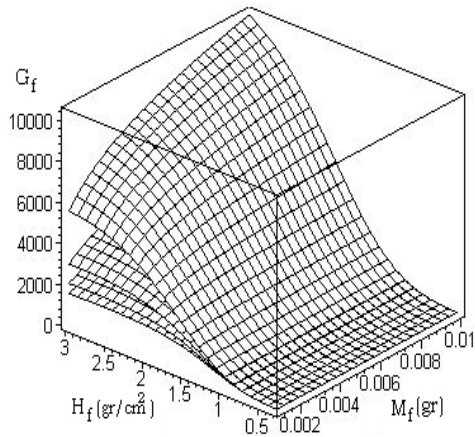

**Fig. 7** The three dimensional variations of $G_f$ versus $M_f$ and $H_f$ for different values of $\alpha$ parameter from top 1,2,3,4.

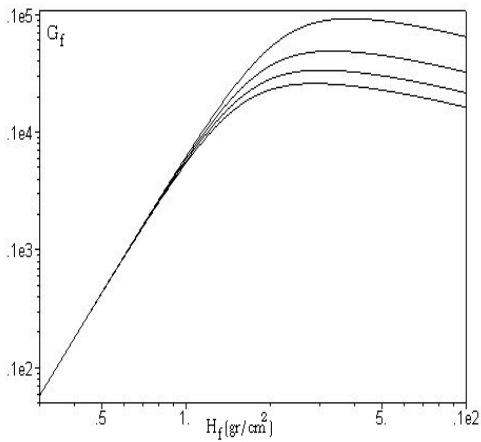 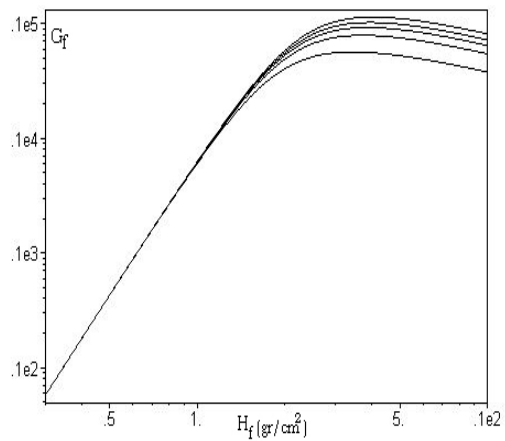

**Fig 8** The variations of fuel gain versus $H_f$ with $M_f=1$ mgr and $\alpha$ From top 1,2,3,4.

**Fig 9** The variations of fuel gain versus $H_f$ with $\alpha=1$ and $M_f$ from top 10,7,5,3,1 mgr.

## 3. Results

Generally in mentioned central spark ignition method, we want the pellet design and the related parameters with that to be the way which fuel gain is higher in value. Therefore according to equation (4), $E_f$ must have the maximum value and consequently according to equation (10), the value of $H_f$ have to increase. Also the denominator of equation (4) must be the minimum value and therefore according to equation (14) and (17), a maximum limit for $H_f$ exists. This value that for it either $E_f$ has higher value or $E_c+E_s$ has the least value, is the desired limit for design of pellet and fuel density inside it, which figures 6 and 7 show it clearly in different physical conditions. Therefore according to these diagrams, the best value for $H_f$ is yielded too. Gain diagrams show that for increasing $H_f$ value from 0.3 gr/cm$^2$ to 1 gr/cm$^2$, the value of fuel gain increases and this increase is independent on fuel mass and isentrope parameter. For increasing of $H_f$ from 1 gr/cm$^2$, fuel gain increases as for more fuel mass and less isentrope parameter, fuel gain has a bigger value. This condition continues to $H_f = 3$ gr/cm$^2$ and after that fuel gain decreases. Consequently the value of $H_f$ equal to 3 gr/cm$^2$ is optimized value for pellet design.

# Appendix

We know that the 14.1 MeV neutrons from d-t nuclear fusion can release this energy in fusion fuel region, due to elastic collisions. But otherwise due to neutrality of neutrons, they will have leakage from the pellet region. The value of neutrons leakage energy in an isomolar d-t fuel is calculated by the following equation [5]:

$$E = \varepsilon\left(1 - e^{\frac{-\delta \rho R}{m}}\right) \qquad (19)$$

where $\varepsilon, \delta, m, \rho$ and R are average neutron energy decreases in one collision, average elastic scattering cross section, average ion mass, density and fuel pellet radius respectively. Fig. 10 shows the variations of deposited neutron energy versus time with different $H_f$ values.

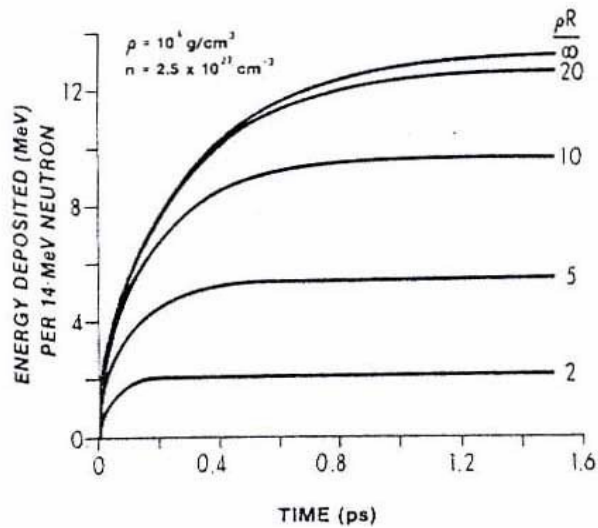

**Fig.10** The variations of neutron energy deposited versus Time for different values of $H_f$ [5]

Fig.10 shows that the results for an equimolar DT sphere at a density of $10^+$g/cm$^3$. It indicates that confinement parameter $\rho R$ values greater than about 5 is necessary. For example, $\rho R = 5$, 30% of the energy is deposited before the neutron leave the medium, or in another case with $\rho R$ greater than 20, neutron energy deposition function is similar. The reason why energy deposition does not approach its full value (14.1 MeV) is that some energy is lost by the d(n,2n)p reaction which absorbs about 2.5MeV. The deposition time scale is relatively small, typically about 0.2 Pico seconds at this density. For comparison, the capsule burn time scale will be larger than former at least an order of magnitude, so that neutron energy deposition is effectively instantaneous. The neutron energy decrease in each collision is [6]:

$$n = \frac{1}{\zeta} Ln\left(\frac{E_0}{E}\right) \quad , \quad \zeta \approx \frac{2}{\frac{2}{3} + A_{eff}} \tag{20}$$

where $n, \zeta, E_0$, $E$ and $A_{eff}$ are collision number, average lethargy increase, neutron energy before collision, neutron energy after collision and mass number of the material in which neutron enters it respectively. With substitution of known values and without considering the equation d(n,2n)p we will have :

$$E = E_0 \exp[-0.64698n] \tag{21}$$

Regarding to Fig.11, it's implied that neutron with initial energy of 14.1 MeV will need exactly seven collision for depositing all of its energy in an infinite, homogeneous and equimolar d-t fusion region.

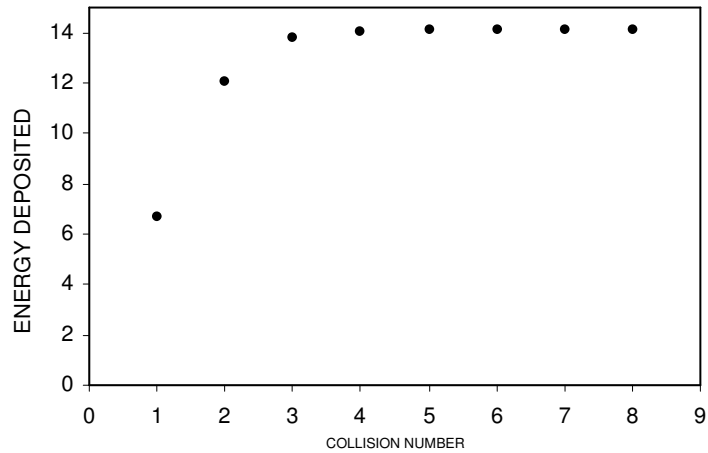

**Fig.11** The variations of neutron energy deposited versus collision number

By consideration fusion neutron energy leakage and equations (5) and (18), our fuel gain decreases to 80%.

**Acknowledgment:**
The authors greatly appreciate stimulating discussion with *S.Khoshbinfar* and guilan university computer centre.

**References**
[1]  A.A.Harms, et al. "Principles of fusion energy", world scientific pub.(2000)
[2]  R.E.Kidder, Nuclear Fusion, 16, 3 (1976)
[3]  M.Tabak, Nuclear Fusion, 36, 2 (1996)
[4]  J. Meyer-ter-vehn, Nuclear Fusion, 22, 4 (1982)
[5]  S. Skupsky, et al. , nuclear fusion, 18, 6 (1978)
[6]  J. R. Lamarsh, "Introduction to nuclear engineering", Addison Wesley pub.(1983)